# Electrical spin injection from ferromagnet into an InAs heterostructures through MgO tunnel barrier

T. Ishikura, Z. Cui, L-K. Liefeith, K. Konishi and Kanji Yoh

Research center for integrated quantum electronics, Hokkaido university,

Sapporo 060-8628, Japan

We have investigated electrical spin injection from $Ni_{81}Fe_{19}$ into an InAs quantum well through MgO tunneling barrier for potential application to InAs-based spin field effect transistor. Injected spin polarized current were detected in both nonlocal and local spin valve set-ups and the spin diffusion length and spin injection efficiency were analyzed. The spin diffusion length was estimated to be 1.6 μm in nonlocal set-up at 1.4 K. The spin polarization of $Ni_{81}Fe_{19}$/MgO/InAs as-deposited sample was relealed to be 6.9 %, while increased spin polarization of 8.9 % was observed by additional thermal treatment.

Semiconductor spintronics has a potential of providing alternative logic scheme to overcome a limitation of conventional Si VLSI technology.[1-4] For improving the Spin-field effect transistor (FET) characteristics, high spin injection efficiency and strong spin orbit (SO) interactions are desirable. So far a number of spin injection and detection in ferromagnetic metal/semiconductor hybrid structures have been examined in various methods, such as circular polarization detection from spin light emitting diode[5-6], electrical spin accumulation detection in nonlocal geometry[7-9] and so on. Among them, nonlocal spin valve measurement has been widely used for electrical spin injection and detection measurements which is free from the influence of anisotropic magnetic resistance (AMR) and local Hall effect. Moreover, conductivity mismatch at the interface of FM and semiconductors (SC) is one of the obstacles for suppressing spin injection efficiency in case of zero electric field at the interface.[10] This can be overcome by using either Schottky barrier or tunneling barrier near FMs/SC interface, which have been investigated in various FM/SC materials[11]. Optimized control of the contact resistance plays a crucial role for efficient spin injection and detection. In those circumstances, InAs heterostructure is a good candidate to investigate the spin injection dependence on the contact resistance by changing the tunneling barrier thickness because it is well known that the Fermi level is pinned inside conduction band. On the other hand, for gate control of the channel spin state is another issue to achieve spin transistor. For this purpose, InAs is known to be a good candidate material for spin channel in a Datta-Das spin transistor for its high spin orbit interaction[12-15]. The band offset and carrier distribution was found to affect the Rashba SOI strength strongly and gate electric field in InAlAs/InGaAs metamorphic heterostructures make it possible to control it.[16-20] Thus, the FM/MgO/InAs structure has been selectedas a suitable materials system in order to investigate the advantages of MgO tunnel barrier on spin injection from FM into narrow gap semiconductors. In this letter, we show local and nonlocal spin injection from NiFe spin injector into an InAs heterostructures through MgO tunnel barrier and investigated the spin diffusion length and spin injection efficiency dependence on various structures.

We have fabricated the local and nonlocal spin injection devices based on an InAs-based inverted high electron mobility transisistor (HEMT) structure grown on semi-insulating InP (001) substrate. The 4.1 nm InAs channel layer of the HEMT is buried in the middle of InGaAs sub-channel layer to confirm mobility and increased SO interaction. The structure consist of 300/800 nm of InP/ $In_{0.52}Al_{0.48}As$ buffer layer, $1.2 \times 10^{12}$ cm$^{-2}$ of Si δ-doping, 10 nm of $In_{0.52}Al_{0.48}As$ spacer layer, 5.6 nm of $In_{0.53}Ga_{0.47}As$ sub-channel, 4.1 nm of InAs main channel, 1.8 nm of $In_{0.53}Ga_{0.47}As$ sub-channel, 50 nm of $In_{0.52}Al_{0.48}As$ barrier layer and $In_{0.53}Ga_{0.47}As$ 10 nm cap-layer. The electrical transport parameters were characterized by Hall measurement, yielding 20,000 cm$^2$/Vs of mobility, $1.5 \times 10^{12}$ cm$^{-2}$ of carrier density and 300 Ohm/□ of sheet resistance at 1.4 K. The SO interaction strength was observed to $23 \times 10^{-12}$ eVm at 1.4 K determined by Shubnikov-de Haas oscillation and its waveform analysis by Maximum Entropy Method of Fast Fourier Transform (FFT). The channel of 50 μm width was constructed by wet chemical etching aligned along [110] direction. The FM electrode patterns for a spin injector and a detector were defined to be 0.2 and 0.4 μm widths using electron beam lithography. The $Ni_{81}Fe_{19}$ (70 nm)/MgO (2 nm) stack was deposited by electron beam evaporator on InAs channel directly after wet chemical etching of top barrier layers. After lift-off, a part of the sample was subjected to annealing at 250 °C for 40 min in ultra high vacuum chamber for the purpose of MgO crystalline quality improvement. Then, the contact resistance•area product (RA) of the ferromagnetic electrode to the InAs channel was controlled by MgO thickness and annealing treatment. The measured RA without anneal were $2 \times 10^{-5}$ Ωcm$^2$ and ≈$3 \times 10^{-4}$ Ωcm$^2$ for non-annealed and with annealed samples characterized by 4-terminal measurement method, indicating that the annealing process reduced leakage current. The increase of contact resistance may have been caused by decrease of anti-site defects in MgO by the thermal treatment. These resistance values are sufficient for achieving appreciable spin injection based on conductivity mismatch theory $R^s_{ch}=R_sL_{sd}/W$ where $R_s$ is the sheet resistance, $L_{sd}$ is the spin diffusion length and W is the channel width.[21]

Spin valve measurements were conducted in local and nonlocal configurations at 1.4 K as shown in Figure 1. The nonlocal voltage was measured between electrodes 3 and 4, whereas the constant current of 100 µA was run between terminals 2 and 1. The magnetic field was swept parallel and anti-parallel to the longitudinal direction of NiFe electrodes. In Figure 2, typical nonlocal (a) and local (b) measurement results are shown. The spacing of terminals 2 and 3 of the nonlocal measurement was 0.7µm. The clear voltage peaks were obtained at ≈100 Gauss and at ≈ -100 Gauss upon positive and negative sweeps of the external magnetic field, respectively.

Additionally, local measurement results are shown in Fig.2 (b). Its resistance peaks are seen to coincide with the nonlocal measurement shown in Fig.2 (a). The magnetoresistance (MR) ratio was estimated to be 0.2 %, which is 6 times higher than nonlocal voltage signals. In Fig.3, the $R_{NL}$ ($V_{NL}/I$) and MR ($|R_{AP}-R_P|/R_P$) ratio are shown as a function of L. Spin polarization at the injection point, η, and spin diffusion length, $L_{sd}$, were estimated by fitting the data with the formula $R_{NL}=(\eta_I\eta_D R_s L_{sd}/W)\exp(L/L_{sd})$, where η is the spin polarization of spin injector and detector and $L_{sd}$ is spin diffusion length. The estimated spin polarization and spin diffusion length for the non-annealed sample were 6.9 % and 1.6 µm, respectively and annealed sample, they were 8.9 % and 0.73 µm, respectively. To the best of our knowledge, it is the highest electrical spin injection from ferromagnet into InAs channel. Additionally, the spin life time $\tau_S$ was estimated to be 8.0 ps from $\tau_S=L_{sd}^2/D$, where diffusion constant, D, was substituted by the extracted value from Hall-effect measurements assuming that it is independent of spin polarization. Unfortunately, it is not possible to compare this estimation of $\tau_S$=8.0 ps with Hanle measurements because the Hanle signal of the present sample was below detection level within 1k Gauss scan. This result seems reasonable because estimated magnetic field of the Hanle half-width would amount to 4 T if we use $\tau_S$=8.0 ps value and the formula $\tau_S$=h/(2πg$\mu_B$B), where h is Plank constant, g is effective electron g-factor, $\mu_B$ is Bohr magnetron and B is perpendicular magnetic field against substrate. Thus, it is obvious that we can never obtain $\tau_S$ with Hanle measurement without changing the magnetization directions from in-plane to perpendicular direction because 4 T is twice as high as vertical saturation magnetization (hard axis) of the ferromagnetic electrode

The MR magnitude is theoretically estimated from contact resistance. When the contact resistance is dominated by tunnel resistance, MR ratio is described as follows.[22]

$$MR = \frac{\eta^2}{1-\eta^2} \frac{2}{2\cosh\left(\frac{L_{ch}}{L_{sd}}\right) + \left(\frac{R_b}{R^s_{ch}} + \frac{R^s_{ch}}{R_b}\right)\sinh\left(\frac{L_{ch}}{L_{sd}}\right)}. \quad (1)$$

, where $R_b$=$R_P$/(2(1-$\eta^2$)). This relationship predicts maximum MR signal when $R_b$ equals to $R^s_{ch}$, the normalized channel resistance. Following equation (1), MR dependence on $R_b$ of the non-annealed and annealed samples are plotted in Fig.4 based upon nonlocal spin valve measurement results. For the non-annealed sample, our experimental data exhibited an MR ratio of 0.2 % with a contact resistivity of 2×10$^{-5}$ Ωcm$^2$ and $L_{ch}$=700 nm, whereas the annealed sample showed an MR ratio below 0.02 % with a contact resistivity of 3×10$^{-4}$ Ωcm$^2$ and $L_{ch}$=1μm. This explains why local measurement signal of the annealed sample was below detection level. Both MR ratios are in good agreement with the previous calculation. Both improvement of spin injection efficiency (8.9%) and increased tunneling resistance (3x10$^{-4}$Ωcm$^2$) of the annealed sample suggest that the annealing process improved the crystalline quality of e-beam evaporated MgO tunnel layer.

In summary, we have demonstrated spin injection efficiency of 8.9 % in NiFe/MgO/InAs structure in nonlocal spin valves measurements. Non-annealed samples showed spin injection efficiency of 6.9 % and 5.7 % by nonlocal and local measurements, respectively. The enhancement of spin injection efficiency from 6.9 % to 8.9 % by annealing is presumably due to the improvement of e-beam evaporated MgO crystal quality upon annealing. Measured value of local MR agreed reasonably well with the estimation based on nonlocal measurements.

Acknowledgement

This work was partly supported by Strategic International Research Cooperative Program, Japan Science and Technology Agency (JST).

References

[1]S. Datta and B. Das, Appl. Phys. Lett. **56,** 665 (1990).

[2]P. Bruno and J. Wunderlich, J. Appl. Phys. **84,** 978 (1998).

[3]A. F. Morpurgo, J. P. Heida, T. M. Klapwijk, B. J. van Wees and G. Borghs, Phys. Rev. Lett. **80** 1050 (1998).

[4]S. Sugahara and M. Tanaka, Appl. Phys. Lett. **84**, 2307 (2004).

[5]H. J. Zhu, M. Ramsteiner, H. Kostial, M. Wassermeier, H. P. Schonherr, and K.H. Ploog, Phys. Rev. Lett. **87**, 016601 (2001).

[6]K. Yoh, H. Ohno, K. Sueoka, and M. E. Ramsteiner, J. Vac. Sci. Technol. B **22**, 1432 (2004).

[7]H. C. Koo, H. Yi, J. B. Ko, J. Chang, S. H Han, D. Jung, S. G. Huh and J. Eom, Appl. Phys. Lett **90**, 022101 (2007).

[8]X. Lou, C. Adelmann, S. A. Crooker, E. S. Garlid, J. Zhang, K. S. M. Reddy, S. D. Flexner, C. J. Palmstrom and P. A. Crowell, Nature Physics **31**, 197 (2007).

[9]T. Sasaki, T. Oikawa, T. Suzuki, M. Shiraishi, Y. Suzuki, and K. Tagami, Appl. Phys. Express **2**, 053003 (2009).

[10]G. Schmidt, D. Ferrand, L. W. Mollenkamp, A. T. Fillip and B. J. van Wees, Phys. Rev. B **62**, R4790 (2000).

[11]E. I. Rashba, Phys Rev, B **62**, R16 267 (2000).

[12]I. Žutić, J. Fabian and S. D. Sarma, Rev. Mod. Phys. **76**, 323 (2004).

[13]J. Schliemann, J. C. Egues and D. Ross, Phys. Rev. Lett. **90**, 146801 (2003).

[14]R. Winkler, Phys. Rev. B **69**, 045317 (2004).

[15]M. Ohno and K. Yoh, Phys. Rev. B **75,** 241308(R) (2007).

[16]P. Pfeffer and W. Zawadzki, Phys. Rev. B **68**, 035315 (2003).

[17]J. Nitta, T. Akazaki, H. Takayanagi, and T. Enoki, Phys. Rev. Lett. **78**, 1335 (1997).

[18]T. Matsuda and K. Yoh, Physica E **42**, 979 (2010).

[19]T. Matsuda and K. Yoh, J. Electron. Mater. **37**, 1806 (2008).

[20]H. C. Koo, J. H. Kwon, J. Eom, J. Chang, S. H. Han and M. Johnson, Science **325,** 1515 (2009).

[21]F. Magnus, S. K. Clowes, A. M. Gilbertson, W. R. Branford, E. D. Barkhoudarov, L. F. Cohena, L. J. Singh, Z. H. Barber, M. G. Blamire, P. D. Buckle, L. Buckle, T. Ashley, D. A. Eustace and D. W. McComb, Appl. Phys. Lett. **91**, 122106 (2007).

[22]B. Dlubak, M. B. Martin, C. Deranlot, B. Servet, S. Xavier, R. Mattana, M. Sprinkle, C. Berger, W. De Heer, F. Petroff, A. Anane1, P. Seneor and A. Fert, Nature Physics **8**, 557 (2012).

Figure captions

FIG. 1. Schematic diagram of four-terminal lateral spin valves. The wide channel of 50μm was formed by wet chemical etching. After the top barrier layers were etched down to the InAs channel layer, NiFe/MgO were stacked. The width of the spin injector and detector (NiFe/MgO) were to be 0.2 and 0.4 μm. For spin valve measurements, the magnetic field was swept in long distance (1-10) direction of the electrodes. Nonlocal voltage was detected between 3 and 4, while constant current was biased form 2 to 1.  The local resistance was detected between 2 and 3.

FIG. 2. Typical spin valve characteristics in nonlocal geometry (a) and local geometry (b) in L=0.7 μm at 1.4 K. The current was fixed to 100 μA for nonlocal and 1 mA for local measurements. The closed circle (open circle) represents positive (negative) direction sweep of the external magnetic field along the longitudinal axis of FM electrodes.

FIG. 3. The spin valve signals as a function of FM electrode spacing, L, in nonlocal geometry (a) and local geometry (b) at 1.4 K. The dotted line indicates the exponential decay fit. (c) The spin valve signals of the annealed samples and its exponential decay fit. ($L_{sd}$=1.6 μm)

FIG. 4. The MR ratio as a function of contact resistance of the non-annealed (square) and annealed samples (circle). The values of theoretical calculation were indicated by solid line for non-annealed and dotted line for annealed samples, respectively.

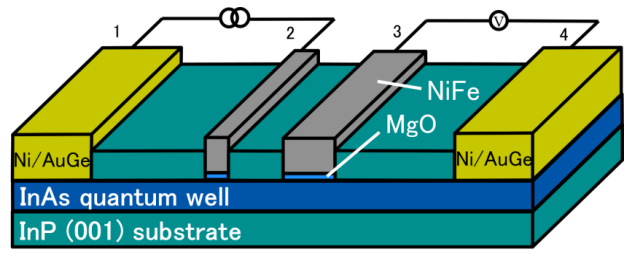

FIG. 1

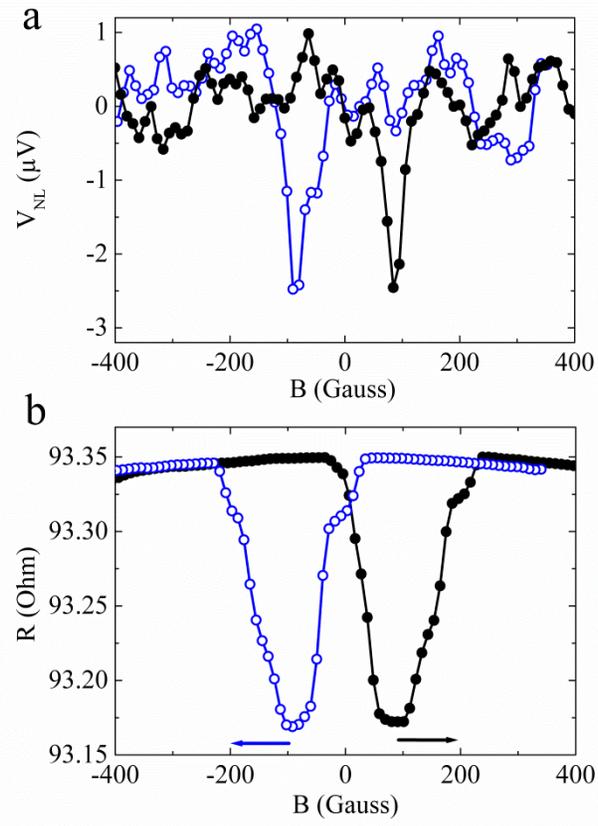

FIG. 2

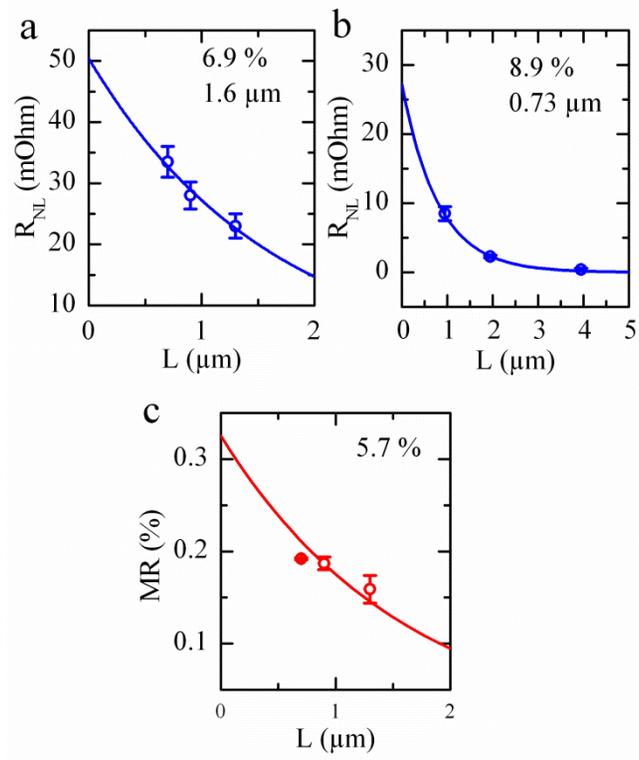

FIG. 3

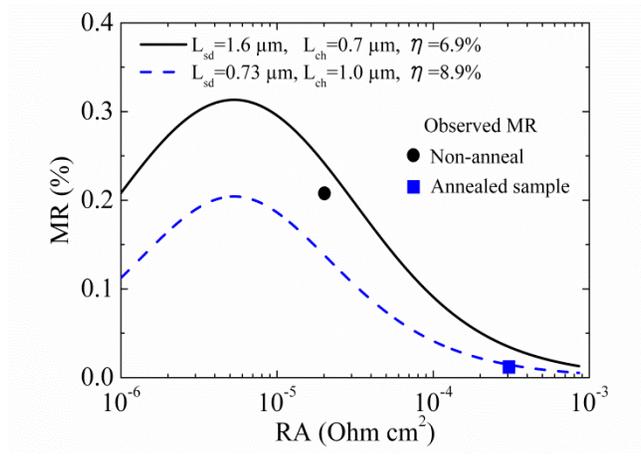

FIG. 4